\documentclass[fleqn,twoside]{article}
\usepackage[headings]{espcrc2}

\readRCS
$Id: espcrc2.tex,v 1.2 2004/02/24 11:22:11 spepping Exp $
\usepackage{graphicx}
\usepackage[figuresright]{rotating}

\newcommand{\AmS}{{\protect\the\textfont2
  A\kern-.1667em\lower.5ex\hbox{M}\kern-.125emS}}

\newcommand{\nn}{\nonumber}
\newcommand{\bea}{\begin{eqnarray}}
\newcommand{\eea}{\end{eqnarray}}

\def\dex{\int_0^1 dx}

\def\ra{\rangle}

\def\spa#1.#2{\langle#1\,#2\rangle}
\def\spb#1.#2{[#1\,#2]}

\def\spab#1.#2.#3{\langle\mskip-1mu{#1}
                  | #2 | {#3}]}

\def\spba#1.#2.#3{[\mskip-1mu{#1}
                  | #2 | {#3}\rangle}

\def\spbb#1.#2.#3.#4{[\mskip-1mu{#1}
                     | {#2} \ {#3} | {#4}]}

\def\spaa#1.#2.#3.#4{\langle\mskip-1mu{#1}
                     | {#2} \ {#3} | {#4}\rangle}

\def\dea{\langle \ell \ d \ell \rangle}
\def\deb{[\ell \ d \ell]}

\newbox\charbox
\newbox\slabox
\def\s#1{{      
        \setbox\charbox=\hbox{$#1$}
        \setbox\slabox=\hbox{$/$}
        \dimen\charbox=\ht\slabox
        \advance\dimen\charbox by -\dp\slabox
        \advance\dimen\charbox by -\ht\charbox
        \advance\dimen\charbox by \dp\charbox
        \divide\dimen\charbox by 2
        \raise-\dimen\charbox\hbox to \wd\charbox{\hss/\hss}
        \llap{$#1$}
}}

\title{On the Cuts of Scattering Amplitudes}

\author{Pierpaolo Mastrolia\address{CERN PH-TH, 
        CH-1211 Geneva 23, Switzerland}%
        \thanks{email: {\tt Pierpaolo.Mastrolia@cern.ch }}}
       
\begin{document}

\begin{abstract}
The use of complex analysis for computing one-loop
scattering amplitudes is naturally induced by generalised 
unitarity-cut conditions, fulfilled
by complex values of the loop variable. 
We report on two techniques: the cut-integration with spinor-variables
as contour integrals of rational functions; 
and the use of the Discrete Fourier Transform to optimize
the reduction of tensor-integrals to master scalar integrals.   
\vspace{1pc}
\end{abstract}

\maketitle

\pagestyle{empty}
\thispagestyle{empty}

\section{GENERALISED UNITARITY}

The application of unitarity as an on-shell method of 
calculation \cite{Bern:1994zx,BernMorgan} is based on 
the principles that  products of on-shell tree-level amplitudes
produce functions with the correct branch cuts in
all channels; and that any one-loop amplitude is expected
to be expressed, by Passarino-Veltman reduction, as 
a linear combination of scalar master integrals (MI's),
that are characterised by their own, leading and subleading,
singularities \cite{Bern:2007dw}.

Dimensionally-regulated amplitudes can be decomposed in terms
of MI's in shifted-dimensions as,
\bea
\mathcal{A}_N^{(4-2\epsilon)} = 
\sum_{n, \ 0 \le j \le j_{\rm max}}
 c_{nj} \times \mathcal{M}_n^{(4-2\epsilon+2j)} \ ,
\label{eq:DdimDeco}
\eea
where $\mathcal{A}_N^{(4-2\epsilon)}$ is any $N$-point 
amplitude in $D$-dimensions (being $D=4-2\epsilon$),
and $\mathcal{M}_n^{(4-2\epsilon+2j)}$ is a $n$-point MI's in 
$(4-2\epsilon+2j)$-dimensions,
with $n\in\{0,2,3,4,\ldots,N\}$, and $j_{\rm max}$
depending on the process.
In Eq.(\ref{eq:DdimDeco}),
the coefficients $c_{nj}$ do not depend on $D$, and 
the whole $D$-dependence is embedded in the definition of 
$\mathcal{M}$'s~\cite{BernChalmers,BernMorgan},
\bea
&&\hspace*{-0.7cm}\mathcal{M}_n^{(4-2\epsilon+2j)}
= 
{\Omega_{-1-2\epsilon}
\over
\Omega_{-1-2\epsilon+2j}
} \ 
(4 \pi)^{j} 
I_n^{(4-2\epsilon+2j)} \ , \\
&&\hspace*{-0.7cm}
\Omega_{k} = 
{2\pi^{k+1 \over 2} \ 
\Gamma^{-1}({(k+1) / 2}) } \ ,
\eea
where $\Omega_{k}$ is the generalised solid-angle,
and $I_n^{(4-2\epsilon+2j)}$ is the scalar $n$-point
function in shifted dimensions.
The decomposition (\ref{eq:DdimDeco}) could be
further simplified with the help of recurrence
relations linking higher-point integrals
to lower-point ones \cite{Bern:1992em}.
From Eq.(\ref{eq:DdimDeco}), it is clear that
the computation of the amplitude requires the
knowledge of two types of ingredients: the
MI's, and their coefficients.
In the following, we focus on the determination of the latter ones.

The principle of a unitarity-based method is the
extraction of the {\it rational} coefficients, $c_{nj}$, by
matching the multiparticle cuts of the amplitude onto
the corresponding cuts of the MI's.
By considering \cite{Mahlon:LoopMeasure,BernMorgan,Brandhuber:2005jw} 
the splitting of $L_D$, the loop momentum in 
$(4-2\epsilon)$-dimension, into its four-dimensional
component, $L$, and its orthogonal complement $L_{(-2\epsilon)}$,
as $L_D \equiv L + L_{(-2\epsilon)}$ 
(with
$L_D^2  \equiv L^2 + L_{(-2\epsilon)}^2 $ and 
$L_{(-2\epsilon)}^2 \equiv - \mu^2$
),
the $D$-dimensional integration measure can be written
as a convolution of a four-dimensional integration and
an integration on $\mu^2$,
\bea
\int d^D L_D &=& 
\int d^{-2 \epsilon} \mu \int d^4 L \ = \\
&=& \Omega_{-2\epsilon} \int d\mu^2 \ (\mu^2)^{-1-\epsilon} \int d^4 L \ .
\eea
By applying the above splitting to both sides of Eq.(\ref{eq:DdimDeco}),
the four-dimensional kernel of the
amplitude, ${\cal A}_N^{(4)}$, 
can be read as expressed 
in terms of four-dimensional $n$-point MI's, $I_n^{(4)}$, 
\bea
{\cal A}_N^{(4)}(\mu^2) = 
\sum_{n,j} \ c_{nj} \times (\mu^2)^j \times I_n^{(4)}(\mu^2) \ .
\label{eq:4dimDeco}
\eea
We notice that the $\mu^2$-dependence of ${\cal A}_N^{(4)}$
is due to the presence of $\mu^2$ 
in all the denominators of $I_n^{(4)}$, as additional mass-term, and
to the polynomial coefficients, $c_{nj} \times (\mu^2)^j$.
It is therefore evident that to find out $c_{nj}$, it is sufficient
to compute the $\mu^2$-polynomials \cite{Britto:2008sw} that are the
coefficients of the four-dimensional MI's, $I_n^{(4)}$,
in the decomposition  (\ref{eq:4dimDeco}).

$I_n^{(4)}$ are functions determined by their own branch-cuts.
Generalised unitarity is a very effective tool to extract the
rational coefficients of functions by exploiting
their singularity structure, which is accessed
by imposing (on-shellness) cut-condition to propagating
particles, 
\bea
(q^2 - m^2 + i0)^{-1} \to \ (2\pi i) \ \delta^{(+)}(q^2 - m^2) \ .
\eea
In general, the fulfillment of multiple-cut conditions requires
loop momenta with complex components.
Since the loop momentum, $L$, in Eq.(\ref{eq:4dimDeco}) 
has four components, the effect of the cut-conditions is to
freeze some of its components, when not all, according to the
number of the cuts. With the {\it quadruple}-cut \cite{Britto:2004nc}
the loop momentum is completly frozen, yielding the
algebraic determination of the coefficients 
of $I_n^{(4)}, (n\ge4)$;
the spinorial integration of the
{\it double}-cut \cite{Britto:2005ha,Britto:2006sj,ABFKM,FordeTriBub}
and {\it triple}-cut \cite{FordeTriBub,MastroliaTriple,BjerrumBohr:2007vu}
lead to the reconstruction of $I_2^{(4)}$- and 
$I_3^{(4)}$-coefficients;
while the coefficients of $I_0^{(4)}$ are detected by {\it single}-cut.
In cases where fewer than four denominators are cut, the loop momentum
is not frozen: the free-components are left over as integration variables 
of the phase-space. 
We will discuss two strategies for dealing with the degrees of freedom
represented by those variables: {\it i)} analytic integration
of the phase-space with spinor variables
\cite{Britto:2005ha,Britto:2006sj,ABFKM,Britto:2007tt,Britto:2008vq,Britto:2008sw}; {\it ii)} algebraic
decomposition of the integrand by means of the Discrete Fourier Transform
(DFT)~\cite{Mastrolia:2008jb}.


\section{DOUBLE-CUT AND SPINORS}

By using the splitting of the 
loop variables as above,
the $D$-dimensional double-cut in the $P$-channel of
any amplitude, {\it i.e.} the
double-cut of the {\it l.h.s}
of Eq.(\ref{eq:DdimDeco}), can
be written as a convolution
of a four-dimensional double-cut
and an integration on $\mu^2$,
\bea
\Delta(\mathcal{A}_N^{(D)}) = 
\int d \mu^{-2\epsilon}
\ \Delta({\cal A}_N^{(4)}) \ ,
\eea
where
$\Delta({\cal A}_N^{(4)})$ is the
double-cut of the {\it l.h.s}
of Eq.(\ref{eq:4dimDeco}),
\bea
&&\hspace*{-0.7cm}
\Delta({\cal A}_N^{(4)}) = 
\int \! d^4 L \ 
\delta^{(\!+\!)}\!(L^2 \!-\! M_1^2 \!-\! \mu^2) \nn \\ &&
\times \delta^{(\!+\!)}\!((L \!-\! P)^2 \!-\! M_2^2 \!-\! \mu^2) 
\ A_1^{\rm tree} \ A_2^{\rm tree} \ ,
\eea
with $A_i^{\rm tree}$ being the tree-amplitudes sewn in the cut.
We found it convenient to decompose \cite{ABFKM} the
four-dimensional loop variable, $L$,
in terms of a massless momentum $\ell$,
and the momentum accross the cut, $P$,
\bea
&& \hspace*{-0.7cm}L_\nu = t \ \ell_\nu + z_0 \ P_\nu \ , \\
&& \hspace*{-0.7cm}t = (1-2z_0) P^2/\spab \ell.P.\ell \ ,\\
\label{eq:z0def}
&& \hspace*{-0.7cm}z_0 = 
     (P^2+M_1^2-M_2^2-\sqrt{\lambda - 4\mu^2})/2P^2 \ , \\
&& \hspace*{-0.7cm}\lambda  =  
(P^2)^2 \!+\! (M_1^2)^2 \!+\! (M_2^2)^2 \!+\! \nn \\ &&
\!-\! 2 P^2 M_1^2 
\!-\! 2 P^2 M_2^2 
\!-\! 2 M_1^2 M_2^2 \ , 
\eea
with $z_0$ being the anomalous threshold, and $\lambda$, the K\"allen
function.
With the above transformation,
$\Delta({\cal A}_N^{(4)})$ can be written in terms of spinor-variables
\cite{Britto:2005ha,Britto:2006sj},
$|\ell\rangle$ and $|\ell]$
(associated to the massless momenta, $\ell$, through
$\s{\ell} = |\ell\rangle[\ell|$), 
and
can be cast as a sum of terms whose general
structure reads,
\bea
&& \hspace*{-0.7cm}\Delta({\cal A}_N^{(4)}) = \sum_i \Delta_i \ , 
\quad \Delta_i =
\int \dea \deb \ {\cal I}_i \ , \\
&& \hspace*{-0.7cm}{\cal I}_i = 
\rho_i\left(|\ell\rangle \right)
{ 
\spb \eta.\ell^{n}
\over 
\spab \ell.P_1.\ell^{n+1}
\spab \ell.P_2.\ell
}
\ ,
\label{pm:eq:4Dgendeco}
\eea
where $P_1$ and $P_2$ can either be equal to the cut-momentum $P$, 
or be a linear combination of external vectors;
and where the $\rho_i$'s depend solely on one spinor flavour, say 
$|\ell\rangle$ 
(and not on $|\ell]$), and may contain poles in $|\ell\rangle$.
We give as understood the dependence of ${\cal I}_i$ on $\mu^2$,
through the variable $z_0$.
The explicit form of the vectors $P_1$ and $P_2$ 
in (\ref{pm:eq:4Dgendeco}) is determining 
the nature of the double-cut, logarithmic or not,
and correspondingly the topology of the 
diagram which is associated to:
if $P_1 = P_2 =P$,
\bea
{\cal I}_i =
\rho_i\left( |\ell\rangle \right)
{ 
\spb \eta.\ell^{n}
/ 
\spab \ell.P.\ell^{n+2} 
} 
\ , \quad
\label{pm:eq:InoFpar}
\eea
and the result will be non-logarithmic,
hence corresponding to the cut of
a 2-point function with external momentum $P$;
if $P_1 = P$, $P_2 \ne P$ or $P_1 \ne P_2 \ne P$,
one proceeds by introducing
a Feynman parameter, to write ${\cal I}_i$ as,
\bea
&& \hspace*{-0.7cm}
{\cal I}_i = 
(n+1) \dex \ 
(1-x)^n \ 
{ 
\rho_i\left( |\ell\rangle \right)
\spb \eta.\ell^{n}
\over 
\spab \ell.R.\ell^{n+2} 
} 
\ , \quad
\label{pm:eq:IwithFpar} \\
&&\hspace*{-0.7cm}
\s{R} = x \s{P}_1 + (1-x) \s{P}_2 \ ,
\label{pm:eq:Rdef}
\eea
and (because of the parametric integral)
the result is logarithmic, hence containing 
the cut of a linear combination of 
$n$-point functions with $n\ge3$.
The spinorial structure of Eq.~(\ref{pm:eq:InoFpar}) and 
Eq.~(\ref{pm:eq:IwithFpar}) is the same. Therefore,
we discuss the spinor integration of the latter,
because it is more general.

\subsection{Contour Integrals}

One can proceed with a change of variables \cite{BrittoHP2},
decomposing $|\ell\rangle$ and $|\ell]$
into two arbitrary massless momenta, $p$ and $q$
(light-cone decomposition),
\bea
&& \hspace*{-0.7cm}
\forall p,q: q^2 = p^2 = 0 \ , \\
&& \hspace*{-0.7cm}
|\ell\ra \equiv |p\rangle + {z} |q\rangle \quad 
|\ell] \equiv |p] + {\bar{z}} |q] 
\label{eq:ChangeVars} \\
&& \hspace*{-0.7cm} 
\dea \deb = - \spab q.p.q \ dz \ d\bar{z} \ .
\eea
Its effect on $\Delta_i$ reads,
\bea
&& \hspace*{-0.7cm} \Delta_i = 
(n+1) \dex \ 
(1-x)^n \ 
\spab q.p.q \nn \\
&& \times
\oint dz \ d\bar{z} \ \rho_i( { z} ) 
{ 
(\spb \eta.p + {\bar{z}} \spb \eta.q)^n
\over 
\chi^{n+2}(z,\bar{z})
} \ , \\
&& \hspace*{-0.7cm} \chi(z,\bar{z}) = 
\spab p.R.p
+ {z}\spab q.R.p +  \nn \\ && \quad
+ { \bar{z}}\spab p.R.q
+ {z} { \bar{z}} \spab q.R.q \ .
\eea
One may observe that
the $z$-$\bar{z}$-integrand 
can be written as a total derivative
with respect to $\bar{z}$
\bea
&& \hspace*{-0.7cm}\Delta_i = 
 \dex \ 
(1-x)^n \ 
\spab q.p.q \nn \\
&& \times
\oint dz \ d\bar{z} 
 {d \over d {\bar{z}}} \bigg\{
\rho_i(z)
{ 
(\spb \eta.p + {\bar{z}} \spb \eta.q)^{n+1}
\over 
\xi(z) \ 
\chi^{n+1}(z,\bar{z})
}\bigg\}
\eea
with
$\xi(z) = (\spab p.P.q + z \spab q.P.q)$,
so that the spinor integration
has been turned into a contour integral of
a rational function in $z$,
\bea
&& \hspace*{-0.7cm} \Delta_i = 
 \dex \ 
(1-x)^n \ 
\spab q.p.q \nn \\
&& \times
\oint dz 
\bigg\{
\rho_i(z)
{ 
(\spb \eta.p + {\bar{z}} \spb \eta.q)^{n+1}
\over 
\xi(z) \ 
\chi^{n+1}(z,\bar{z})
}\bigg\} \ .
\label{pm:ImplicitDoubleCut}
\eea
The $z$-integral can be performed
by Cauchy's residue theorem, summing
the residues at the poles in $z$
(substituting as well $\bar{z} = z^{*}$).
There are two sources of poles to account for:
{\it i)} the poles contained in $\rho_i(z)$;
{\it ii)} the pole due to $\xi(z)$, whose
value is 
\bea
z_\xi=-\spab p.P.q/\spab q.P.q \ .
\label{eq:zxipole}
\eea

To complete the integration of $\Delta_i$ in (\ref{pm:ImplicitDoubleCut}),
one has to perform the parametric integration which is finally
responsible for the appearence of logarithmic terms in the double-cut. 
On the contrary, the spinorial integration in (\ref{pm:eq:InoFpar})
would generate a contribution without branch-cuts.
We remark that the role of $|\ell\rangle$ and $|\ell]$ in the integration
could be interchanged.

At the end of the phase-space integration, by adding up all the 
$\Delta_i$'s, one finally gets a result whose structure is
\bea
\Delta({\cal A}_N^{(4)}) = 
\sum_{2 \le n,j} \ c_{nj} \times (\mu^2)^j \times \Delta(I_n^{(4)}) \ ,
\label{eq:4dimCutDeco}
\eea
corresponding to the double-cut of Eq.(\ref{eq:4dimDeco}).
Out of (\ref{eq:4dimCutDeco}), it is possible to extract the 
polynomial coefficients, $c_{nj} \times (\mu^2)^j \ (n \ge 2)$: 
the coefficient of $I_0^{(4)}$, the tadpoles,
cannot be detected within the double-cut, and their
determination should be provided by independent
informations on the amplitude.
We recall that the $\mu^2$-dependence of the coefficients
originates from the understood presence of $z_0$, given in Eq.(\ref{eq:z0def}).

We observe that a proper choice of 
the momenta $p$ and $q$, entering the change of variables
(\ref{eq:ChangeVars}), 
can simplify dramatically the calculation.
For instance, 
they determine the value of the $z_\xi$-pole, given in Eq.(\ref{eq:zxipole}): 
given $q_\mu$, and the cut-momentum $P_\mu$, 
the choice $p_\mu \equiv P_\mu - q_\mu \times P^2/\spab q.P.q$, 
would yield $z_\xi=0$.

The phase-space integration just discussed
was used succesfully for an analytic computation
of non-trivial one-loop corrections.
In particular, its four-dimensional (massless) version
\cite{Britto:2005ha,Britto:2006sj,MastroliaTriple}
has been applied to complete the non-supersymmetric 
cut-constructible term 
of the six-gluon amplitude in QCD \cite{Britto:2006sj}, 
to compute the six-photon amplitude in QED \cite{Binoth:2006hk,Binoth:2007ca}, 
and the cut-constructible term of 
a general MHV amplitudes involving a Higgs plus $n$-gluons
in QCD (in the heavy-top limit) \cite{Glover:2008ff}.

Recently, the efficiency of spinor
integration has been pushed to achieve
closed analytic forms for the
generating formulas of the coefficients
of $I_n^{(4)} \ (2 \le n \le 4)$, for an arbitrary massive
process \cite{Britto:2008vq,Britto:2008sw}, 
which together with $I_0^{(4)}$ constitute
a basis of functions in four-dimensions,
hence, due to the relation among Eq.(\ref{eq:4dimDeco}) and 
Eq.(\ref{eq:DdimDeco}), in $D$-dimensions.
The formulas presented in \cite{Britto:2008vq,Britto:2008sw} 
- too long to be shown here -
can be evaluated, without performing any integration,
by specializing the value of input variables that are specific to the 
initial cut-integrand as assembled from tree-level amplitudes.

We have as well recently released the package {\tt S@M} 
(Spinors @ Mathematica)\cite{Maitre:2007jq}, that implements the 
spinor-helicity formalism in Mathematica. 
The package allows the use of complex-spinor algebra 
along with the multi-purpose features of Mathematica, and it is suitable
for the algebraic manipulation and integration 
of products of tree amplitudes with
complex spinors sewn in generalised unitarity-cuts.

\section{OPTIMIZED REDUCTION}


\newcommand{\beq}{\begin{equation}}
\newcommand{\eeq}{\end{equation}}
\newcommand{\bqa}{\begin{eqnarray}}
\newcommand{\eqa}{\end{eqnarray}}
\newcommand{\nl}{\nonumber \\}
\def\db#1{\bar D_{#1}}
\def\d#1{D_{#1}}
\def\tld#1{\tilde {#1}}
\def\slh#1{\rlap / {#1}}
\def\eqn#1{Eq.~(\ref{#1})}
\def\eqns#1#2{Eqs.~(\ref{#1}) and~(\ref{#2})}
\def\eqnss#1#2{Eqs.~(\ref{#1})-(\ref{#2})}
\def\fig#1{Fig.~{\ref{#1}}}
\def\figs#1#2{Figs.~\ref{#1} and~\ref{#2}}
\def\sec#1{Section~{\ref{#1}}}
\def\app#1{Appendix~\ref{#1}}
\def\tab#1{Table~\ref{#1}}


As an alternative to any phase-space integration, 
in \cite{Ossola:2006us,Ossola:2007bb} there was proposed
a very efficient method for the reconstruction 
of the coefficients in the decomposition (\ref{eq:4dimDeco}).
In what follows, I limit the discussion to the
so called cut-constructible term of a scattering amplitude,
that corresponds to the poly-logarithmic 
structure arising when Eq.(\ref{eq:4dimDeco}) is evaluated
at $\mu^2=0$.
I will sketch the reconstruction of the complete 
$\mu^2$-dependence \cite{Pittausimple,Ellis:2007br,Giele:2008ve,Ossola:2008xq}
at the end of the section.
The by-now known as {\tt OPP}-reduction allows the numerical reconstruction of $c_{n0}$, by solving a system of algebraic equations that are obtained by: {\it i)} the numerical evaluation of the {\it integrand} at explicit values of the loop-variable, on the one side; 
{\it ii)} and the knowledge of the most general {\it polynomial} structure of the {\it integrand} itself~\cite{intlevel}, on the other one. 
The values of the loop momentum used for the numerical evaluation of the integrand
are chosen among the set of solutions of the multiple-cut conditions, 
{\it i.e.} the solutions of the system of
equations obtained by imposing the vanishing of the 
cut-denominators.

\subsection{{\tt OPP}-Reduction}

The starting point of the {\tt OPP} reduction method~\cite{Ossola:2006us,Ossola:2007bb}
is the general expression for the {\it integrand} of a generic $m$-point 
one-loop amplitude that can be written as
\bqa
\label{eq:1}
&&\hspace*{-0.7cm}A_m(q)= 
\frac{N(q)}{\d{0}\d{1}\cdots \d{m-1}}\,,\\
&&\hspace*{-0.7cm}\d{i} = ({q} + p_i)^2-m_i^2\,,~~~ p_0 \ne 0\, ,
\eea
where $N(q)$ is the four-dimensional numerator of the amplitude.\footnote{
$A_m(q)$ is the integrand of ${\cal A}_m^{(4)}(\mu^2=0)$, 
defined in Eq.(\ref{eq:4dimDeco})} 
The main formula of the {\tt OPP}-algorithm is
the expression of $N(q)$ in terms of the denominators
$\d{i}$,
\bea
\label{eq:2}
N(q) &=& \sum_{\alpha=1}^4 \Delta_\alpha(q)
\eea
with
\bea
\Delta_4(q) \!\!\! &=& \!\!\!\!\!\! 
\sum_{i < j < k < \ell}^{m-1}
\left[
          d_{ijk\ell} +
     \tld{d}_{ijk\ell}(q)
\right]
\prod_{\beta \ne ijk\ell}^{m-1} \d{\beta} , \\
\Delta_3(q) \!\!\! &=& \!\!\!\! 
\sum_{i < j < k }^{m-1}
\left[
          c_{ijk} +
     \tld{c}_{ijk}(q)
\right]
\prod_{\beta \ne ijk}^{m-1} \d{\beta} , \\
\Delta_2(q) \!\!\! &=& \!\!\!\! 
\sum_{i < j }^{m-1}
\left[
          b_{ij} +
     \tld{b}_{ij}(q)
\right]
\prod_{\beta \ne ij}^{m-1} \d{\beta} , \\
\Delta_1(q) \!\!\! &=& \!\!\!\! 
\sum_{i}^{m-1}
\left[
          a_{i} +
     \tld{a}_{i}(q)
\right]
\prod_{\beta \ne i}^{m-1} \d{\beta} .
\eea
By inserting (\ref{eq:2}) back in (\ref{eq:1}), one exposes
the multi-pole nature of $A_m$.
The coefficients of the multi-pole expansion can be further 
split in two pieces:
a piece that still depends on $q$, parametrized by
$\tld{d},\tld{c},\tld{b},\tld{a}$, that vanishes upon integration,
and a piece that does not depend on $q$, parametrized as $d,c,b,a$.
Such a separation is always possible, as shown in~\cite{Ossola:2006us}, 
and, with this choice, the latter set of coefficients corresponds
to the ensemble of the coefficients of 
$I_n^{(4)}(\mu^2=0) \ , (n \in \{0,2,3,4\})$: $a,b,c,d$ in (\ref{eq:2}) 
correspond respectively to $c_{00},c_{20},c_{30},c_{40}$ in (\ref{eq:4dimDeco}).

\subsection{Top-Down System}

The goal of the algorithm is reduced to the algebraical problem of fitting
the coefficients $d,c,b,a$ by evaluating the function $N(q)$
a sufficient number of times, at different values of $q$,
and then inverting the system. Accordingly, 
let us define the following functions,
\bea
&&\hspace*{-0.7cm}R_{ijk\ell}(q) \equiv
    N(q) 
       \big( \prod_{\beta \ne ijk\ell}^{m-1} \d{\beta} \big)^{-1}\ , \\
&&\hspace*{-0.7cm}R_{ijk}^{\prime}(q) \equiv 
   (N(q) - \Delta_4(q)) 
       \big( \prod_{\beta \ne ijk}^{m-1} \d{\beta} \big)^{-1}\ , \\
&&\hspace*{-0.7cm}R_{ij}^{\prime\prime}(q) \equiv
   (N(q) - \sum_{\alpha=3}^4\Delta_\alpha(q)) 
       \big( \prod_{\beta \ne ij}^{m-1} \d{\beta} \big)^{-1}\ , \\
&&\hspace*{-0.7cm}R_{i}^{\prime\prime\prime}(q) \equiv 
   (N(q) - \sum_{\alpha=2}^4\Delta_\alpha(q)) 
       \big( \prod_{\beta \ne i}^{m-1} \d{\beta} \big)^{-1}\ .
\eea
We as well define as $\{q\}_{ijk\ell}$ the set of the
solutions of $D_i = D_j = D_k = D_\ell = 0$.
Having defined our setup,
from Eq.(\ref{eq:2}) we can derive the following
sets of equations:
\bea
\label{eq:OPPsystem4} 
&&\hspace*{-0.7cm}\Big[
R_{ijk\ell}(q)  = 
         d_{ijk\ell} +
     \tld{d}_{ijk\ell}(q) 
\Big]_{q\in\{q\}_{ijk\ell}} \ , \\
\label{eq:OPPsystem3} 
&&\hspace*{-0.7cm}\Big[
R_{ijk}^{\prime}(q)  = 
         c_{ijk} +
     \tld{c}_{ijk}(q) 
\Big]_{q\in\{q\}_{ijk}} \ , \\
\label{eq:OPPsystem2} 
&&\hspace*{-0.7cm}\Big[
R_{ij}^{\prime\prime}(q)  = 
         b_{ij} +
     \tld{b}_{ij}(q) 
\Big]_{q\in\{q\}_{ij}} \ , \\
\label{eq:OPPsystem1} 
&&\hspace*{-0.7cm}\Big[
R_{i}^{\prime\prime\prime}(q)  = 
         a_{i} +
     \tld{a}_{i}(q) 
\Big]_{q\in\{q\}_{i}} \ ,
\eea
which {\it must} be solved necessarily
in cascade, top-down:
in Eq.(\ref{eq:OPPsystem4}), $N(q)$ is
a known quantity, namely an input of
the algorithm;
but the {\it l.h.s} of each other equation
becomes a known quantity (numerically evaluable), only
after solving the equation which preceeds it.

\subsection{Polynomial Structures and DFT}

An important observation is due.
The {\it r.h.s} of each of the
equations (\ref{eq:OPPsystem4})-(\ref{eq:OPPsystem1})
is a {\it polynomial} function.
Without presenting their explicit expressions 
(see \cite{Mastrolia:2008jb} for the detailed presentation),
the general structure is the following: 
the variables are the components
of $q$ not-frozen by the cut-conditions;
the degree is known;
while the coefficients are the {\it unknowns} to be
determined.
The problem to be tackled is thus a well known mathematical
subject:
{\it polynomial interpolation}.
In order to find out the coefficients of a polynomial,
one can avoid the numerical inversion of a system, which
is a very delicate operation, 
due to the possibility of a vanishing determinant 
in critical kinematic regions.

The Discrete Fourier Transform (DFT) is a very
efficient tool to extract the coefficients of a polynomial,
by evaluating it at special values of the variables
\cite{Britto:2008vq,Mastrolia:2008jb,Berger:2008sj}.
Let us show how it works in the case
of a polynomial of degree $n$ in one variable, $x$, defined as,
\bea
P_n(x) = \sum_{\ell=0}^n \ c_\ell \ x^\ell \ .
\label{eq:GeneralPn}
\eea
At the first step, one
generates the set of discrete values $P_{n,{ k}} \ ({ k}=0,...,n)$,
\bea
P_{n,{ k}} \equiv P_n(x_{ k}) = 
\sum_{\ell=0}^n \ c_\ell \ \rho^\ell \ e^{-2 \pi i {{ k} \over (n+1)} \ell} \ ,
\eea
by sampling $P_n(x)$ at $(n+1)$ equidistant points on the $\rho$-circle,
\bea
 x_{ k} = \rho \ e^{-2 \pi i {{ k} \over (n+1)}} \ .
\eea
At the second step,
using the orthogonality
\bea
\sum_{j=0}^{n-1} 
e^{ 2 \pi i {k \over n} \ j} \ 
e^{-2 \pi i{k^\prime \over n} \ j}
 = n \ \delta_{k k'} \ ,
\eea
one can obtain the coefficient $c_\ell$ simply by {\it projection},
\bea
c_\ell &=& {\rho^{-\ell} \over n+1}
         \sum_{{ k}=0}^n \ P_{n,{ k}} \  e^{2 \pi i {{ k} \over (n+1)} \ell}
         \, .
\eea
In fact, the {\it r.h.s} of Eq.(\ref{eq:OPPsystem1}) is a 
degree-1 polynomial in a single variable,
whose coefficients are easily determined by
the semi-sum and the semi-difference of two numerical values of $R$.
But the {\it r.h.s} of Eqs.(\ref{eq:OPPsystem3})-(\ref{eq:OPPsystem1})
are multivariate polynomials of higher degree.
To find out their coefficients we used a modified 
DFT, that is a Fast Fourier Transform-like algorithm, suitable
to minimize the number of the numerical calls
respectively of $R^{\prime}, R^{\prime\prime}$, and $R^{\prime\prime\prime}$,
being exactly the same as the number of the unknowns,
and to avoid the kinematical singularities emerging at the vanishing
of the circle-radius $\rho$.
In so doing, one can determine all the unknown coefficients,
among which the $O^{\rm th}$-order ones, 
respectively $d_{ijk\ell}, c_{ijk}, b_{ij}, a_{i}$,
correspond to the coefficients of the MI's in four-dimension.

For the reconstruction of the complete $\mu^2$-dependence
of the coefficients in Eq.(\ref{eq:4dimDeco}), 
the decomposition (\ref{eq:2}), must be
slightly extended to account for the presence of $\mu^2$
\cite{Giele:2008ve,Ossola:2008xq}.
The starting point, in this case, is $A_m^{(4)}(\mu^2)$, which contains
a numerator $N(q,\mu^2)$ and denominators
$\db{i} = \d{i} - \mu^2$.
The reduction proceeds exactly as above, with the difference
Eqs.(\ref{eq:OPPsystem4})-(\ref{eq:OPPsystem1})
containing an extra dependence on $\mu^2$. 
Since the $\mu^2$-dependence is still polynomial,
one can use the DFT also in this case, having to deal with
$R, R^{\prime}, R^{\prime\prime}$, and $R^{\prime\prime\prime}$ 
with $\mu^2$ as additional variable \cite{Britto:2008vq}.
The flexibility of the projection-procedure hereby presented
extends its range of applicability to tackle 
the determination of the coefficients of polynomial structures
wherever should this issue occur.
We finally remark that the parametrization of the free (integration) variables 
as complex unitary phases yields as well 
a very effective performance of Cauchy's
residue theorem within the contexts of factorization- and 
unitarity-based methods,
where the on-shellness properties are naturally captured by polar structures 
of complex phases.

\end{document}